\documentclass[iop]{emulateapj}
\slugcomment{Accepted for publication in the Astrophysical Journal}
\usepackage{graphicx}
\usepackage{epstopdf}

\def\etal{\hbox{et al.}$\,$}
\def\OII{\hbox{[O II]}$\,\,$}

\def\Ha{\hbox{H$\alpha$}$\,$}
\def\Hb{\hbox{H$\beta$}$\,$}

\def\Msun{\rm{\hbox{M$_{\odot}$}}}           % Solar masses

\def\24m{\hbox{24~$\micron$}$\,$}
\begin{document}

\title{The IMACS Cluster Building Survey. III. \\The Star Formation Histories of  Field Galaxies\footnotemark[1]}

\footnotetext[1]{This paper includes data gathered with the 6.5 meter Magellan Telescopes located at Las Campanas Observatory, Chile}

 \author{Augustus Oemler Jr\altaffilmark{2}, Alan Dressler\altaffilmark{2}, Michael G. Gladders\altaffilmark{3}, Jacopo Fritz\altaffilmark{4}, Bianca M.\ Poggianti\altaffilmark{5}, Benedetta Vulcani\altaffilmark{5}\altaffilmark{6}, \& Louis Abramson\altaffilmark{3}}

\altaffiltext{2}{The Observatories of the Carnegie Institution for Science, 813 Santa Barbara St., Pasadena, California 91101-1292, oemler@obs.carnegiescience.edu}
\altaffiltext{3}{Department of Astronomy \& Astrophysics, University of Chicago, Chicago, IL 60637} 
\altaffiltext{4}{Sterrenkundig Observatorium, Universiteit Gent, Krijgslaan 281 S9, B-9000 Gent, Belgium}
\altaffiltext{5}{INAF-Osservatorio Astronomico di Padova, vicolo dell'Osservatorio 5, I-35122 Padova, Italy}
\altaffiltext{6}{\,Department of Astronomy, Padova University, Vicolo Osservatorio 3, I-35122 Padova, Italy}

\begin{abstract}
Using data from the IMACS Cluster Building Survey and from nearby galaxy surveys, we examine the evolution of the rate  of star formation in field galaxies from $ z = 0.60$ to the present. Fitting the luminosity function to a standard Schechter form, we find a rapid evolution of $M_B^*$ consistent with that found in other deep surveys; at the present epoch $M_B^*$ is evolving at the rate of $0.38 Gyr^{-1}$, several times faster than the predictions of simple models for the evolution of old, coeval galaxies. The evolution of the distribution of specific star formation rates (SSFR) is also too rapid to explain by such models.  We demonstrate that starbursts cannot, even in principle, explain the evolution of the SSFR distribution. However, the rapid evolution of both $M_B^*$ and the SSFR distribution can be explained if some fraction of galaxies have star formation rates characterized by both short rise and fall times and by an epoch of peak star formation more recent than the majority of galaxies. Although galaxies of every stellar mass up to $1.4\times10^{11}\Msun$ show a range of epochs of peak star formation, the fraction of ``younger'' galaxies  falls from about 40\% at a mass of $4\times10^{10}\Msun$ to zero at a mass of $1.4\times10^{11}\Msun$. The incidence of younger galaxies appears to be insensitive to the density of the local environment; but does depend on group membership: relatively isolated galaxies are much more likely to be young than are group members.

 \end{abstract}

\keywords{galaxies: evolution, star formation, cosmic evolution}

\section{Introduction}

The last 10 years have seen an explosion in the number of studies of the evolution of star formation in the universe. Observations of tens of thousands of high--redshift objects have been obtained, with redshifts determined by either spectroscopic or photometric means, and star formation rates determined from optical spectroscopy, optical, UV, and infrared photometry, as well as X--ray and radio emission. Much of the focus has been on the evolution of the star formation rate density (Lilly \etal 1996, Madau  \etal 1996, see  Cucciati \etal 2012 for a recent summary). The evolution of the star formation rate density is a fundamental cosmological datum, describing the  star formation properties of universe as a whole, but it contains limited information about the behavior of individual galaxies, whose collective histories are responsible for it. However, there  have also been a large number of studies of the distribution of the star formation rates of individual galaxies and their dependence on galaxy properties, environment, and epoch (e.g. Perez-Gonzales \etal 2005, Bauer \etal (2005), Juneau \etal 2005,  Feulner \etal 2005a,b,  Noeske \etal 2007a, Bell \etal 2007, Scoville \etal 2007, Karim \etal 2011). Although there is considerable disagreement about the shape of the star formation rate (SFR) distribution at a given galaxy mass and epoch, the work of the last decade has produced broad agreement about some aspects of the star formation history of galaxies since at least $z = 1.0$. At every epoch the specific star formation rate ($SSFR = SFR/M_{stars}$) decreases with galactic mass, implying that star formation has a more extended history in low than in high mass galaxies. At every mass, the SFR and SSFR decline with cosmic epoch at a rate which is approximately the same for all masses, suggesting that the two dependancies are roughly separable, i.e.  $SFR(M_{gal},t_{uni}) = f(M_{gal}) \times f(t_{uni})$.

 Much of the emphasis in recent papers has been on the history of mass production: how the mass of stars grows with time in particular kinds of galaxies and environments, leading to galaxies of the forms and masses we see today (e.g. Feulner \etal 2005b, Bell \etal 2007, Walcher \etal 2008). However, to understand what is responsible for the mass production, we need to understand the processes which drive the evolution of the star formation rates of individual galaxies. One does not need to go to cosmological depths to discover that, on average, massive galaxies formed the bulk of their stars more rapidly than less massive ones; the correlation of spectral type with star formation history and the (rough) correlation of spectral type with galactic mass have been known for decades. A recent determination using Sloan data has been done by Jimenez \etal (2005). This shift with time of star formation from more to less massive galaxies  is quite often (and quite confusingly) called ``downsizing''.  However, this term is better applied as Cowie \etal (1996) originally intended, to the purported delay in the formation epoch of less massive galaxies or, more precisely, to the delay in the epoch at which star formation peaked.  These two are quite different phenomena, which Neistein \etal (2006), distinguish as {\em archaeological downsizing}, for the mass dependence of star formation timescales, and {\em downsizing in time} for the mass dependence of the epoch of maximum star formation.
 
 Identifying a delay in galaxy formation, whether mass--dependent or not, does require observations at high redshift. In addition to Cowie \etal, claims to see such an effect have been made by  Feulner \etal (2005), and by Noeske \etal (2007b). In all 3 studies the claimed discovery rests on the existence of galaxies with specfic star formation rates too high to be sustained for the age of the universe at the epoch of observation. However, very high star formation rates, and very high values of SSFR, can be produced for short periods by starbursts, and there have been multiple claims that it is starbursts rather than delayed galaxy formation which are responsible for the observed phenomenon. In this paper we shall reexamine these issues, using a sample of field galaxies from the IMACS Cluster Building Survey (ICBS- Oemler \etal 2012, herafter Paper I). The ICBS is a spectroscopic survey of regions surrounding rich intermediate--redshift clusters performed using the IMACS wide--field spectrograph on the Baade Telescope. Although targeted to clusters, the $30\arcmin$ field of IMACS guaranteed that a large fraction of the  objects observed would be foreground and background galaxies, unrelated to the target clusters, and constituting an unbiased sample of the general field population.  We shall show that starbursts cannot be responsible for the observed evolution in SSFR, and that the epoch of peak star formation of a substantial fraction of massive galaxies was delayed by at least several billion years after that of the oldest galaxies.
 
 The organization of this paper is as follows. In section 2 we describe the data sets we have used, which include  IMACS spectroscopy, direct imaging, and (in two of the four fields) Spitzer \24m photometry. We also use nearby surveys to construct local galaxy samples to provide zero points for the redshift evolution seen in the ICBS. In Section 3 we study the evolution of the galaxy populations, including luminosity functions and the evolution of the specific star formation rate. In Section 4 we discuss the results, but delay until a following paper (Gladders \etal 2012, Paper IV) a full analysis of the history of star formation in galaxies. Throughout this and following papers we shall assume cosmological parameters of $H_o=71kms^{-1}Mpc^{-1}$, $\Omega_{matter}=0.27$, $\Omega_{tot}=1.0$.

\section{Data and Models}

\subsection{Data Sets}

We  use a number of data sets in the following analysis. This is intended  to be a study of the evolution of the {\em general field}, by which we mean the totality of the galaxy content of the universe, unbiased by group or cluster membership. Our primary distant sample is drawn from the ICBS (q.v. Paper I). The ICBS survey fields  were selected because the Red-Sequence Cluster Survey technique (RCS- Gladders \& Yee 2000) which we used indicated the presence of a very rich cluster at a suitable redshift ($0.30<z<0.60$); therefore,  these fields do not represent an unbiased sample of the universe.  We correct this by removing galaxies in a redshift interval about each of the one or two clusters in each field whose presence we judge contributed to the RCS detection. We do {\em not} remove the contents of any other groups or clusters sufficiently poor, or sufficiently different in redshift so that they did not contribute to the RCS signal. The results should be a reasonable, even if not perfect, approximation to a fair general field sample. This selection is described in more detail in Paper I.

The ICBS covers 4 fields, each about $30\arcmin$ in diameter, near the celestial equator. In all fields we have deep multicolor imaging in either the SDSS or Johnson system, and very deep imaging in the Gunn r band. In two of the fields we have Spitzer \24m photometry of virtually the entire field to the confusion--limited depth. The entire spectroscopic sample contains 6008 galaxies, of which 4578 comprise our field sample. About 80\% of these are in the redshift range $0.25<z<0.75$. We have medium resolution ($ \sim5\AA$) spectra of most of the sample, but for 1140 objects (946 field galaxies) we only have low dispersion prism (LDP) spectra, which yield lower accuracy redshifts ($\sigma_z \sim 0.01$) and no optical line strengths.

Star formation rates are based on, in order of decreasing preference, (1) optical emission lines plus \24m flux, (2) \24m flux, (3) \Ha flux, (4) \Hb flux, and (5) \OII flux. New calibrations of all these SFR measures are provided in Paper I; although errors in SFR rates increase for the less--preferred measures, all are on the same systematic scale. Stellar masses are calculated from galaxy luminosities and colors using relations of color vs mass--to--light ratio derived from stellar population models, assuming a Salpeter IMF, by a method similar, but not identical to that of Bell \& de Jong (2001). More details can be found in Paper I. All masses referred to in this paper are {\em stellar} rather than dynamical masses.

  We construct two ICBS field samples. The 4-Field sample is drawn from all 4 ICBS fields, to a  depth of $r = 22.5$ and excludes galaxies with  only LDP spectra. We will use this sample whenever the analysis concerns the environment of the galaxies, for which the LDP redshifts are insufficiently precise. Otherwise, we will use the IR-ALL sample, which  is drawn from the two fields, RCS0221 and SDSS0845 that have Spitzer \24m photometry. This sample includes LDP objects, and goes to a depth of $r=23.0$. In all fields, only about one half of all galaxies above the survey limits were observed- the selection of objects being driven by slit mask design- and a small fraction of those observed did not yield a reliable redshift. Therefore, each galaxy has associated with it a weighting which reflects the incompleteness of the catalog at the magnitude and location of the galaxy.  Details of the ICBS data can be found in Paper I.

We shall also make occasional use of the AEGIS sample (Davis \etal 2007). We have converted their SFR's and masses to a Salpeter IMF by multiplying by a factor of 2. For objects with detections at \24m, the AEGIS group have derived SFR's from Spitzer photometry using the calibration of Le Floc'h \etal (2005); we have converted these values to that appropriate for our own calibration of $SFR = f(F_{24})$ (q.v. Paper I). To convert SFR's based on optical lines to our system, we have remeasured a sample of the DEEP2 spectra on which AEGIS is based, and have derived a conversion between the AEGIS values and our own.

For comparisons with the properties of the present--epoch field, we have constructed 2 local samples. The NGP sample include SDSS objects with redshifts $0.035 \le z < 0.045$ in the region of the North Galactic Cap within $11^h \le \alpha < 13^h$ and $15\degr \le \delta < 45\degr$. SFR's are determined from \Ha, and masses are calculated from the observed SDSS colors, using the calibrations in Paper I. This volume has an appreciably lower density than the cosmic mean, and is deficient in high--mass galaxies. We shall use this sample only for analyses concerning the group membership of galaxies. For all other analyses we use the PM2GC catalog (Calvi, Poggianti, \& Vulcani 2011), which covers the volume $0.03 \le z < 0.11$, $10^h \le \alpha < 14^h45^m$, $-0\fdg28 \le \delta < 0\fdg28$. (This sample is drawn from a very narrow strip, making the determination of group membership very uncertain.) Again, SFR's were determined using \Ha, and masses derived from the reported colors. All of the galaxy samples used in this paper are summarized in Table 1. The redshift ranges are those of the subset of the data actually used in this analysis, and the number of galaxies is the number within the given redshift range. Additionally, the PM2GC sample has been limited to galaxies with masses $M_{gal}\ge4\times10^{10}\Msun$.

%Table 1
\begin{deluxetable}{lccc}
\tablecaption{Galaxy Samples}

\tablehead{\colhead{Sample} & \colhead{z} & \colhead{Area (\sq\degr)}  & \colhead{$N_{gal}$}}
\startdata

NGP     &  0.035--0.045  &  46.1 &  1898  \\
PC2MG\tablenotemark{a} & 0.03--0.11  &  39.9 &  1362  \\
IR-ALL  & 0.20--0.60  &  0.32 &  1547  \\
4 Fields & 0.20--0.60  &  0.63 &  1976  \\
AEGIS   &  0.60--1.00  &  0.50 &  2097  \\
\enddata
\tablenotetext{a}{only galaxies of mass $M\ge4\times10^{10}\Msun$}
\end{deluxetable}

Because star formation rates and masses of all samples, both local and distant, have been determined using the same mix of methods, described in Paper I, there should not be any systematic error in derived galaxy properties between the various redshift samples. One potential source of systematic error remains. Since, as we shall see, mean star formation rates increase rapidly with redshift, an incorrect slope of the calibrations of star formation rates vs optical and/or infrared indicators would result in an incorrect absolute rate of evolution of star formation rate. All of our star formation measures are tied to the calibration of SFR vs\24m flux plus optical emission lines of Calzetti \etal (2010). Since this calibration is  tied to the very reliable extinction-corrected hydrogen recombination line measurements of star formation rates, we think it unlikely to be much in error.

\subsection{Galaxy Evolution Models}
We shall compare our results with two sets of simple models for the history of star formation in galaxies. Both sets assume solar metallicity and a Salpeter initial mass function (IMF). The first set have simple exponential decay star formation histories:

\begin{displaymath}
SFR\sim{e^{ - (t - {t_{form}})/\tau }}
\end{displaymath}

Although the instantaneous turn--on of star formation in these models is decidedly non--physical, they have been the most popular models for star formation in galaxies since at least Tinsley (1972). We take the formation epoch, $t_{form} = 0.9 Gyr$, and take values of $\tau$ ranging from 1 to 10 Gyr, as well as $\tau=\infty$, i.e. a constant star formation rate. The galaxy properties for these evolutionary histories were calculated using the Burzual \& Charlot (2003) stellar models.

%Figure 1
\begin{figure}[h]
\vspace{0.5in}
\figurenum{1}
\epsscale{1.1}
\plotone{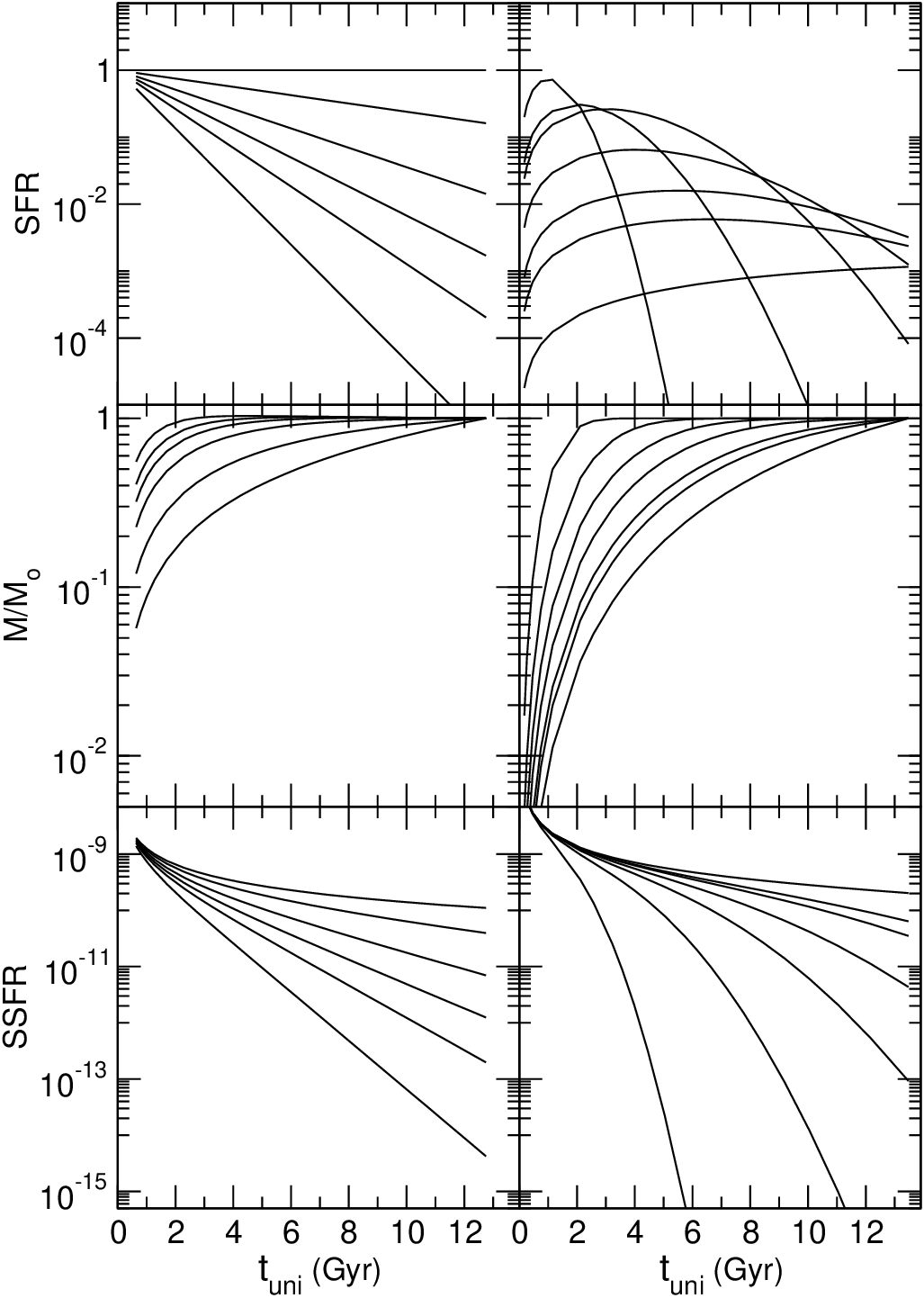}
\caption{ Galaxy evolution models. Left panels- exponential decay models. From bottom to top in the upper and lower panels (top to bottom in the middle panel), models with timescales $\tau$ of 1.0, 1.5, 2.0, 3.0, 5.0, and 7.0 Gyr, and constant star formation. Right panels- delayed exponential models, with timescales, from top to bottom in upper and middle panels, bottom to top in lower panel, $\tau$, of 1.0, 2.0, 3.0, 4.0, 5.5, 6.5, and 12 Gyr. Top panels- evolution of star formation rate, in arbitrary units; middle panels- evolution of stellar mass, in units of the present epoch mass; bottom panels- evolution of specific star formation rate, in units of $Gyr^{-1}$.}
\end{figure}

 Our second set of models assumes star formation histories of the form introduced by Gavazzi \etal (2002):

\begin{displaymath}
SFR\sim t e^{-t^2/2\tau^2}
\end{displaymath}

These are usually  called these {\em delayed exponential} models. For these we calculate galaxy models using a different set of stellar models. These models use the observed stellar libraries of Jacoby \etal (1984) in the optical ($\sim 3400$ to 7400 \AA), and they were extended to the ultraviolet and infrared with the theoretical libraries of Kurucz (1993, private communication). They include emission lines formed in HII regions, that were calculated using the photoionisation code {\sc{cloudy}} (Ferland 1996). The nebular component was calculated assuming {\it case B recombination}, electron temperature of $10^4$ K and electron density of $10^2$ cm$^{-3}$. The source of ionizing photons was assumed to have a radius of 15 pc and a mass of $10^4$ M$_\odot$. The effect of using two different sets of stellar models is negligible compared to the differences due to star formation histories.

The behavior of the two sets of models is presented in Figure 1. Although the star formation histories of the two sets of models are very different, the mass and specific star formation rate histories of the two sets  are quite similar, sharing common characteristics, which will, in fact, be shared by any models with well--behaved (e.g. no starbursts) star formation histories. These include the facts that, at any epoch, galaxies with high SSFR's must have rapidly evolving masses and slowly evolving SSFR's, while galaxies with low SSFR's must have slowly evolving masses and may have rapidly evolving SSFR's. It is also characteristic of these, but not necessarily of every well-behaved model, that the SSFR's within each set of models diverge with time, and never cross. This will be important in our later analysis. 

\section{Evolution of Galaxy Populations}
\subsection{The Luminosity Function}

We determine the Johnson B band luminosity function for our data sample, using a maximum-likelihood fitting of  a Schechter function, as described in Sandage \etal (1979). A preliminary fit to the data, in redshift intervals of 0.1 between $z = 0.13$ and $z=0.73$ yields values for $\alpha$ between -0.90 and -1.35, with no obvious trend with redshift, suggesting that using a fixed value $\alpha = -1.10$ should be suitable for the entire redshift range.  Analysis of other high--redshift galaxy samples have yielded values of $\alpha$ ranging from -1.03 from the zCOSMOS survey (Zucca \etal 2009) to -1.30 from the DEEP+COMBO-17 surveys (Faber \etal 2007), again with no obvious redshift dependance. In Table 2 we present the results of fits to our data, with fixed $\alpha=-1.10$.

%Figure 2
\begin{figure*}
\figurenum{2}
\epsscale{1.0}
\plotone{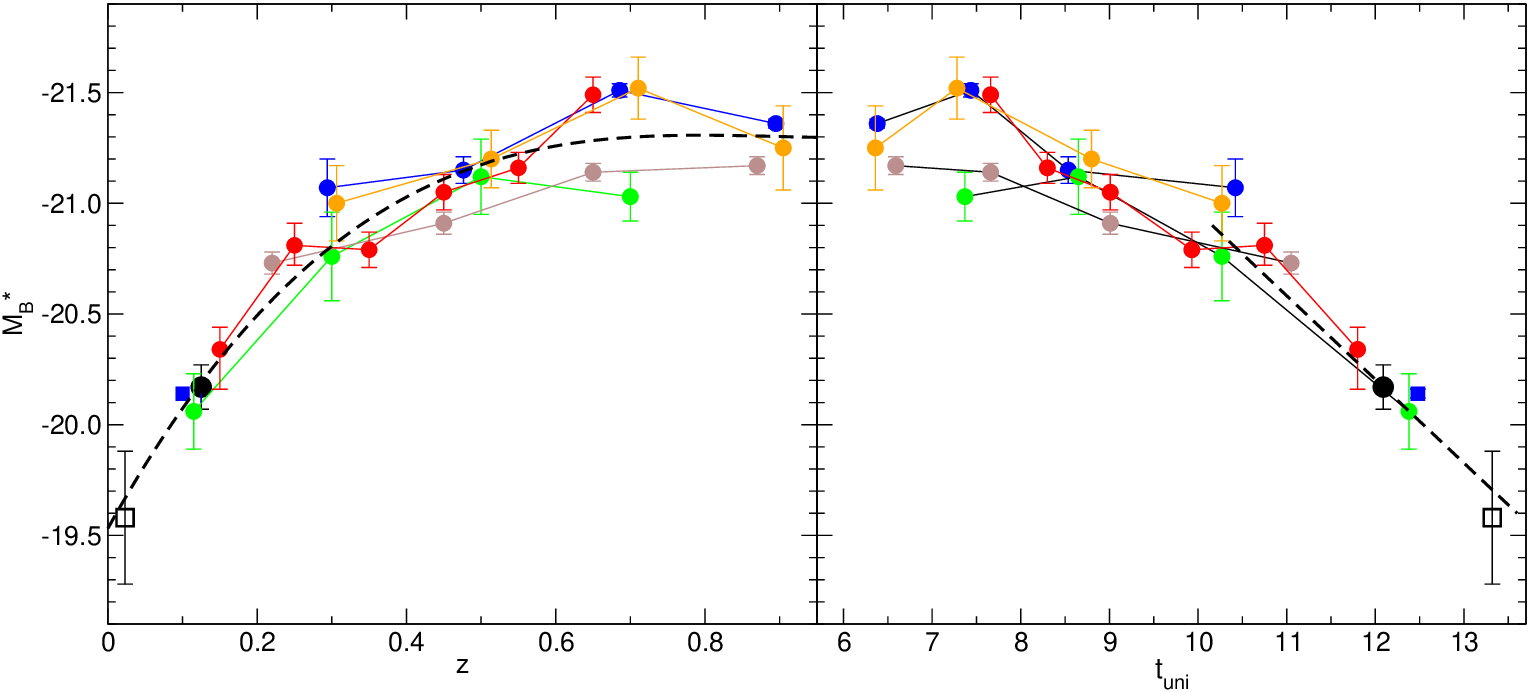}
\figcaption{Evolution of $M_B^*$. Data sets are ICBS- red circles, CfA Redshift Survey- open black square, SDSS- blue square, 2df- black circle, DEEP- blue circles, COBMO-17- orange circles, VIMOS- green circles, zCosmos- brown circles. Dashed lines are trends discussed in text. Left plot is $M_B^*$ vs redshift, right plot is $M_B^*$ vs cosmic epoch.\vspace{0.5cm}}
\end{figure*}

 In Figure 2 we compare the results from Table 2 with other published values at redshifts in the range $0.02 < z < 0.90$, as a function both of redshift and of cosmic epoch. Included are those from the CfA Redshift Survey (Marzke, Huchra, \& Geller 1994),  Sloan Digital Sky Survey (Blanton \etal 2003), 2df survey (Norberg \etal 2002), DEEP and COMBO17 surveys (Faber \etal 2007), VIMOS survey (Ilbert \etal 2005), and ZCosmos survey (Zucca \etal 2009). Although there is a modest scatter between different determinations possibly due to cosmic variance, the overall trend is quite clear: $M_B^*$ has evolved significantly during the last half of the age of the universe, and is evolving most rapidly at the present epoch. 

The composite data are well fit by the relation

\[M_B^* =  - 19.53 - 6.077z + 6.8601{z^2} - 2.5740{z^3}\]

Which is shown by the dashed line in Figure 2a. The dashed line in Figure 2b represents an evolution rate of $M_B^*$ equal to 0.38/Gyr. This is quite rapid, representing a decrease in characteristic luminosity of about a factor of 2 every 2 Gigayears. We can compare this evolution rate with the expectation from the simple models described in the previous section. We take the NGP data set, for each galaxy determine the appropriate value of $\tau$ from its SSFR, and, from the corresponding model calculate $dM_B^*/dt$. We evolve all galaxies back in time for a fixed time interval, fit a Schechter function to the luminosity function at the earlier time, and determine $dM_B^*/dt$. The first three rows of Table 3 compares the observed values of $dM_B^*/dt$, for fixed( $\alpha=-1.10$) and variable values of $\alpha$, with the values predicted by the exponential and delayed exponential models. The predicted evolutionary rates are significantly smaller than the observed rate.

%Table 2
\begin{deluxetable}{cc}
\tablecaption{Evolution of $M_B^*$}
%\tabletypesize[12pt]
%\tablewidth{3in}
\tablehead{\colhead{z} & \colhead{$M_B^*$}}
\startdata
  0.10-0.20	&	$-20.44^{+0.10}_{-0.18} $\\
  0.20-0.30	&	$-20.81^{+0.10}_{-0.09} $\\	
  0.30-0.40	&	$-20.92^{+0.08}_{-0.07} $\\
  0.40-0.50	&	$-21.14^{+0.08}_{-0.08}$ \\
  0.50-0.60	&	$-21.20^{+0.07}_{-0.07} $\\
  0.60-0.70	&	$-21.49^{+0.08}_{-0.08}$\\ 
  \enddata
  \end{deluxetable}

 We have obtained these results for only two quite simple models for galactic evolution, but one can argue that, for any well-behaved model (e.g. no bursts, no special epoch of observation) there is a limit to the evolution rate of $M_B^*$, which depends only of the IMF and the epoch of galaxy formation. If the optical luminosity of a galaxy is dominated by old stars, then, with $t_0\sim13 Gyr$ and a Salpeter IMF,  the evolutionary rate is predicted to be $dM_B^*/dt \sim 0.10/Gyr$. One can obtain a higher rate only if the luminosity is dominated by young stars, and the star formation rate is falling rapidly. However, if the SFR  {\em is} falling fast, then (for well-behaved models) it was much higher in the past, so that the total mass of older stars must be large compared to the mass of young stars, the luminosity  will {\em not} be dominated by the young stars, and the rapid evolution of the SFR will have minimal effect on the evolution of the luminosity. One can escape this trap only by significantly reducing the age of the galaxies, which will reduce the total accumulated number and luminosity of the older stars. In the last two lines of Table 3 we present the predicted evolutionary rates for the delayed exponential models, for two cases in which the start of galaxy formation is delayed for 2 Gyrs and 5 Gyrs. It appears that a decrease of several Gyr in the age of the typical galaxy could explain the observed rapid luminosity evolution.

%Table 3
\begin{deluxetable}{lc}
\tablewidth{0pt}
\tablecaption{Predicted Evolution Rate of $M_B^*$}
\tablehead{\colhead{Sample} & \colhead{$dM_B^*/dt$}}
\startdata
Observed & 0.38 \\
exponential & 0.14/0.24 \\
delayed exponential & 0.10/0.20 \\
delayed exponential + 2 Gyr delay & 0.26/0.32 \\
delayed exponential + 5 Gyr delay & 0.32/0.44 \\
\enddata
\end{deluxetable}

\subsection{Evolution of the Specific Star Formation Rate}

We now turn to the behavior of the specific star formation rates of galaxies. There is, in the literature, some divergence about the shape of the SSFR distribution at a given mass and epoch.  Following Brinchmann \etal (2004), who claim that the  SFR vs mass distribution of nearby galaxies has a quite narrow distribution, Bauer \etal (2005), Noeske \etal (2007a), Peng \etal (2010), and Rodighiero (2011) claim to see a similar ``main sequence'' in the distributions of high--redshift objects. The Brinchmann distribution, as seen in their Figure 17, does have a narrow core among low mass galaxies, but even there the $2\sigma$ tails of the distribution extend out by $\pm$ a factor of $10^2$, and among more massive galaxies with $M_{gal} > 10^{10} \Msun$ even the tight core disappears (see also  Fig. 15 of Salim \etal 2007).  The true breadth of the SSFR distribution can also be seen in the lower redshift bins of Figure 1 of Feulner \etal (2005a). Given the wide range in SFR and SSFR at a given value of Mass, and given that all samples are necessarily incomplete at some level of SFR, the use of a mean value of either quantity seems of limited value. Instead, we will use the normalized cumulative distribution of SSFR, counting from the high end. If we include {\em all} objects in doing the normalization, not just those with detected star formation, this distribution should be valid down to the value of SSFR at which the sample begins to be incomplete.

%Figure 3
\begin{figure*}
\figurenum{3}
\plotone{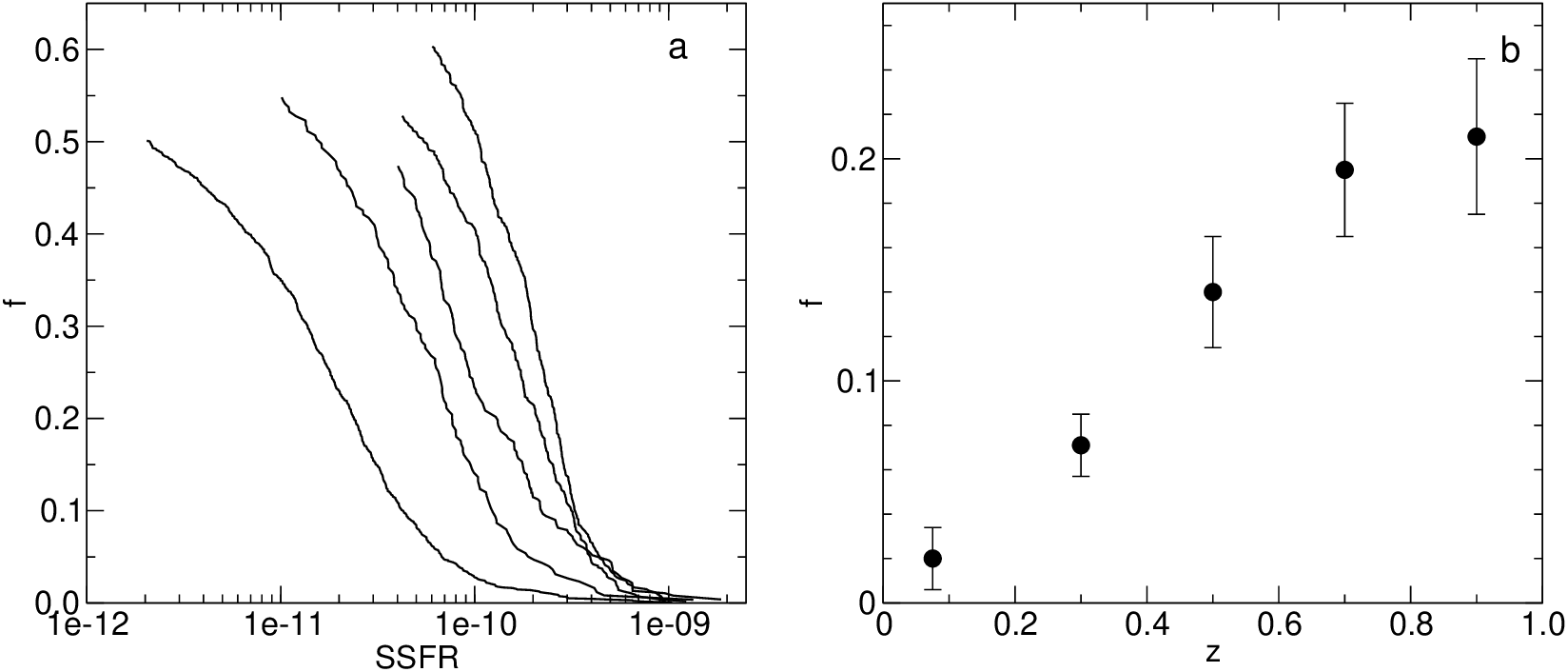}
\figcaption{a- cumulative distribution of specific star formation rate in samples limited by $M_{gal} \ge 4\times10^{10}$. From left to right: PM2GC local sample, ICBS $0.2\le z < 0.4$, ICBS $0.4 \le z < 0.6$, AEGIS $0.6 \le z < 0.8$, AEGIS $0.8 \le z < 1.0$. b- fraction of galaxies with values of SSFR  above the limiting value for a constant star formation rate and Salpeter IMF .}
\end{figure*}

Figure 3a presents the cumulative distribution of SSFR for galaxies more massive than $4\times10^{10}\Msun$,  in 5 redshift intervals, covering $0.0 < z < 1.0$. The lowest redshift group is drawn from the PM2GC sample, and the higher redshift groups are drawn from the IR-ALL and AEGIS samples. As has been noted many times before, the distribution of SSFR moves steadily to higher values  with increasing redshift.  The models summarized in Figure 1 all predict an increase of SSFR with lookback time, but with a rate of change which decreases with SSFR. They also predict a rather hard upper limit to the observed SSFR at any epoch, unless the SFR is increasing rapidly, as in a burst. These are expected to be quite general conclusions, for reasons very similar to those given in the previous section for the evolution of $M_B$. The total stellar mass of a galaxy is

\[{M_{tot}} = \int_{{t_{form}}}^{{t_0}} {SFR(t)fdt} \]

The quantity $f$ is the fraction of the mass that remains in stars, rather than returning to the ISM during stellar evolution. The value of $f$ is about 0.7 for a Salpeter IMF, and somewhat smaller for some other IMFs. Then, at any epoch the observed specific star formation rate

\[\begin{array}{rcl}
SSFR(t) &  = & SFR({t})/\int\limits_{{t_{form}}}^{{t}} {SFR(t)fdt} \\
        &  = & SFR({t})/f(t - {t_{form}})\left\langle {SFR} \right\rangle 
\end{array}\]

If the SFR is not increasing with time, then the maximum value of SSFR will occur when $SFR(t_0) = \langle SFR \rangle$, meaning constant SFR, and will be

\[SSF{R_{\max }} = \frac{1}{{f({t} - {f_{form}})}}\]

For a Salpeter IMF and an early formation epoch $SSF{R_{\max }}$ will be about $1\times10^{-10}$. Furthermore, the evolution rate of $SSFR_{max}$ will be

\[d(\log (SSFR))/dt = 0.4343\frac{1}{{f({t} - {t_{form}})}}\]

For the same parameters, $d(\log (SSFR))/dt$ will today ($t = t_0$) be about 0.05/Gyr. Higher evolutionary rates are only possible for values of  $SFR(t_0)<\langle SFR\rangle$, i.e. for galaxies with declining rates of star formation, and with $SSFR(t_0)<\langle SSFR\rangle$.

Figure 3b presents the fraction of galaxies in the samples in Figure 3a with values of specific star formation higher than the limiting value expected for constant star formation rate and  a Salpeter IMF. This fraction increases from about 2\% for local galaxy samples to 21\% by $z \sim 0.9$ Furthermore, Fig. 3a shows that the shape of the distributions of SSFR changes very little, at least out to a redshift of 0.6, contrary to the expectation that the highest SSFR galaxies should evolve less rapidly than those with lower SSFRs.

\subsection{Are Starbursts Responsible for the Rapid Evolution of Star Formation?}

Observations of both the luminosity function and the specific star formation rate at earlier epochs show that these quantities have evolved faster than expectation. Several mechanisms have been proposed to explain this high evolution rate. The most common suggestion has been starbursts.  Beginning with Dressler \& Gunn (1983) and Couch and Sharples (1987), starbursts have often been invoked to drive galaxy evolution, particularly in clusters (see also Bekki 1999, Poggianti \etal 1999, Miller \etal 2006, Oemler \etal 2009), but also in field galaxies (Le Fevre \etal 2000, Bridge \etal 2007, Bell \etal 2005, Bauer \etal (2005), Dressler \etal 2009).

The main dissent has come from the AEGIS team (Noeske \etal 2007a),  and, at higher redshift, from Rodighiero \etal (2011), which have argued that because the SFR vs mass relation is narrow and well-defined at all redshifts,  only a small number of galaxies could be undergoing a large burst of star formation at any time. Given the breadth of the SFR vs mass distribution, this test is not particularly sensitive to smaller bursts. Nevertheless, one can make a very general argument that, although a starburst can substantially change the instantaneous star formation rate in a galaxy,  starbursts cannot have a significant effect on the distribution of specific star formation rates of a population of galaxies. In Figure 4 we present  the results from several simulations of the effect of adding starbursts to a galaxy population. We make the extreme assumption that 100\% of galaxies have starbursts, either with an amplitude of 3 times the average SFR and a duty cycle of 1/5, or an amplitude of 7 times the average SFR  and a duty cycle of 1/10.  Neither produce a significant enhancement of the high end of the SSFR distributions. The reason is simple: starbursts cannot, by definition, change the time averaged SFR. Periods of enhanced SFR must be balanced by periods of depressed SFR. For any reasonably shaped SSFR distribution, the latter effect offsets almost completely the former.

This is a sufficiently important point that it is worth repeating another way. One can contrive any history of star formation that one wants, driven by starbursts or any other conceivable process. However, the {\em specific} star formation history is not so easily changed, since it is a ratio which depends in its denominator on the entire past history of star formation. Unless one is observing a special epoch, before which starbursts did not occur, adding bursts of star formation to a galaxy's history will also add {\em mass} and therefore will not by itself change the time--averaged specific star formation rate. Stated yet another way, the SSFR measures the characteristic (inverse) {\em timescale} for star formation, not the star formation rate.

%Figure 4

\begin{figure}[h]
\vspace{0.4in}
\figurenum{4}
\plotone{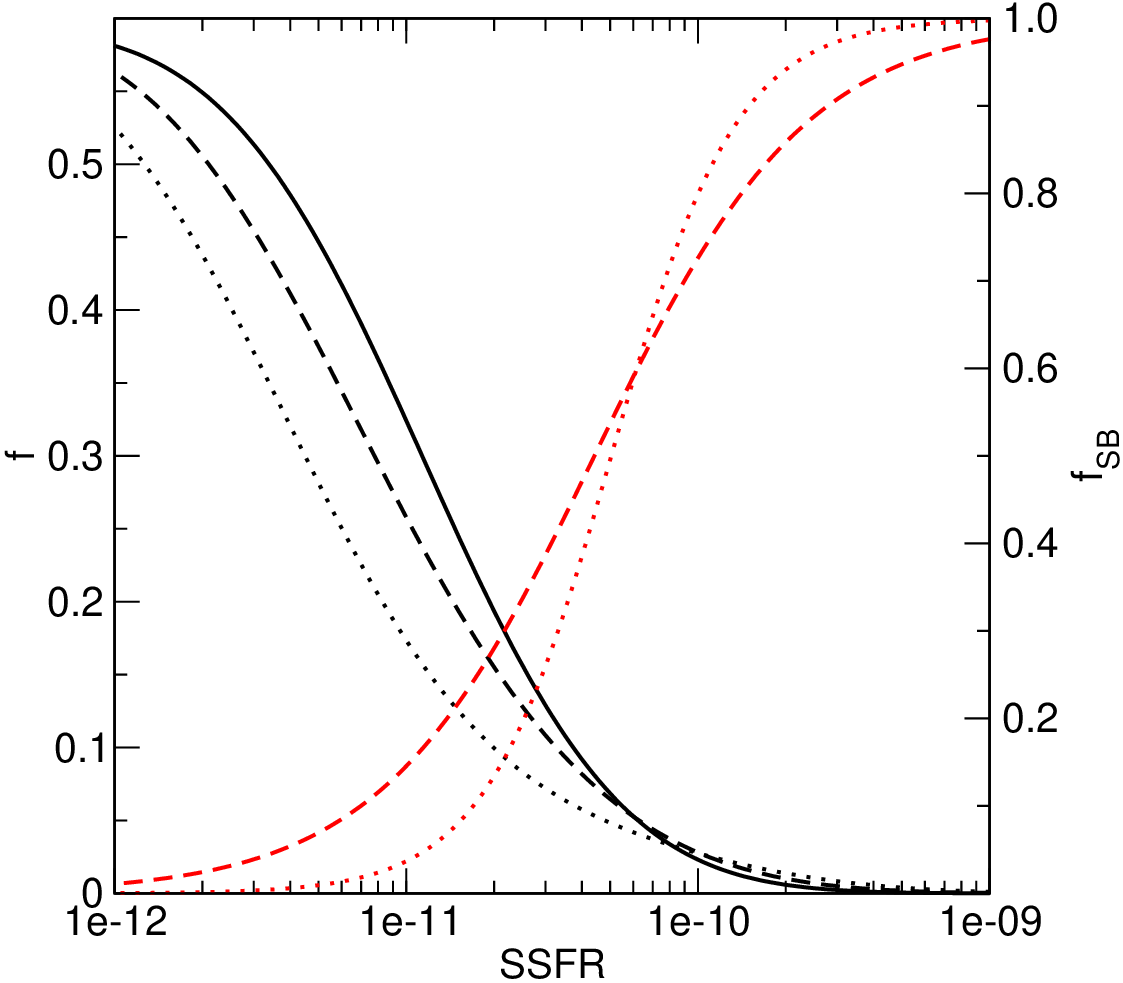}
\figcaption{The black lines, with the scale at the left of the graph, show the effect of starbursts on the distribution of SSFR in a galaxy population. Solid line-  Local SSFR; dashed line- population with time averaged SFR of the Local population but galaxies have 3:1 starbursts with a duty cycle of 1/5; dotted line- population with 7:1 starbursts and 1/10 duty cycle. The red lines, with the right--hand scale, show the fraction of objects in the two starburst scenarios, which are observed in the starburst phase, as a function of SSFR.}
\end{figure}

%Figure 5
\begin{figure}[h]
\figurenum{5}
\epsscale{1.1}
\plotone{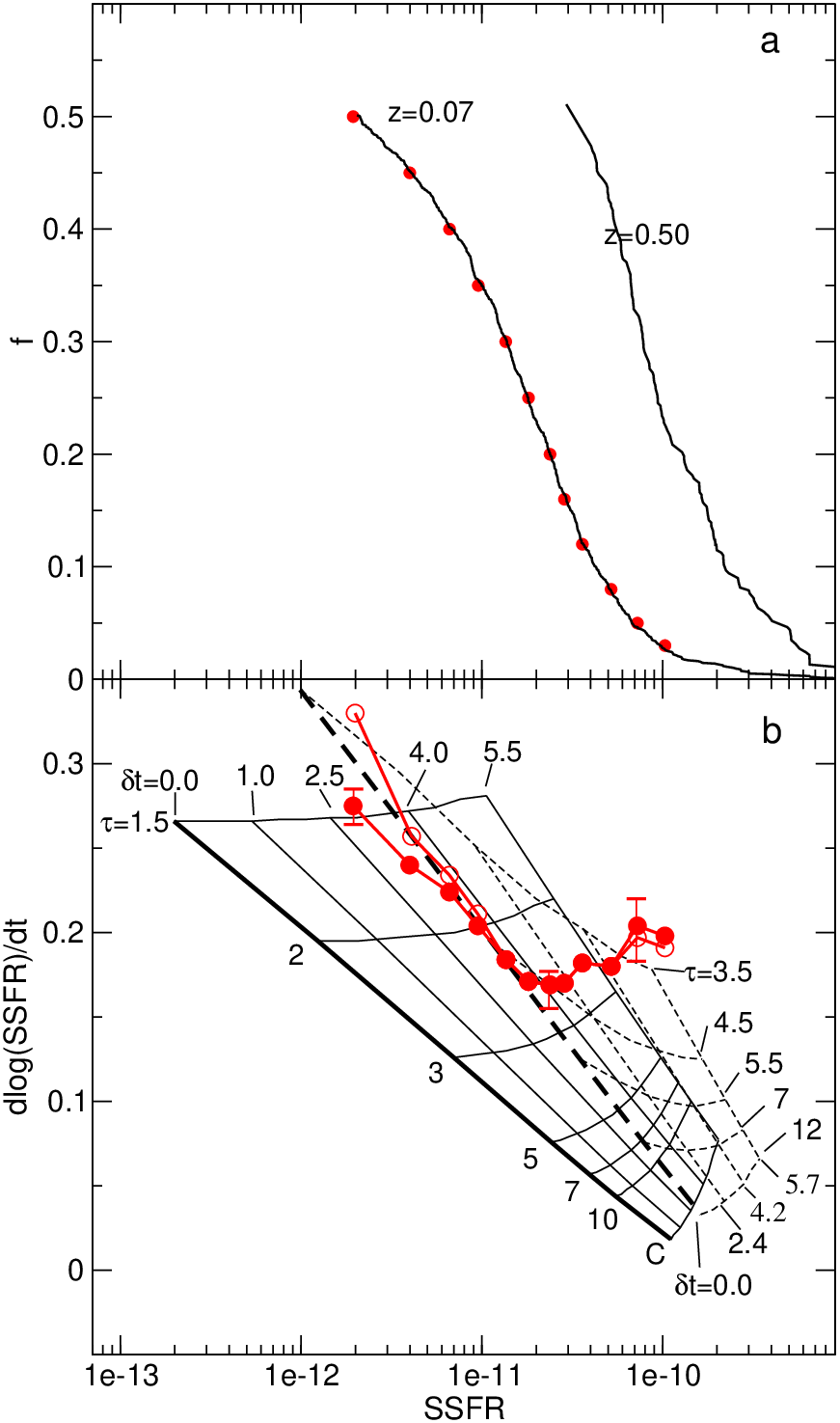}
\figcaption {a- The cumulative distribution of specific star formation rate in samples limited by $M_{gal}<4\times10^{10}$. left- PM2GC sample, right- IR-ALL sample with $0.40 \le z <0.60$. b- evolutionary rate of SSFR vs $SSFR(t_o)$. The filled red points correspond to the values of f shown as red points on the local population curve in the top plot. Open red circles are corrected for the effect of mergers, as described in the text. The two grids correspond to exponential (solid) and delayed exponential (dashed) galaxy evolution models with indicated values of tau, and indicated delays of the start of star formation (in Gyr).}
\end{figure}

This is not to say, however, that starbursts, if common, cannot effect the SSFR values of individual galaxies. The red curves in Fig. 4 present, for the two starburst scenarios, the fraction of objects at a given value of SSFR that are observed to be in a starburst. Starbursts do not have a large net effect of the SSFR distribution because the starburst cycle moves objects to lower as well as to higher values. However, those with the highest values at any time are those currently in a starburst phase. Although they do not emphasize the fact, Bell \etal (2005) find a factor of two higher merger rate among galaxies with the highest values of SSFR; and, Bridge (2005) find that, at $0.2 \le z < 1.3$, galaxies with detected \24m flux are a factor of 5 more more likely to have close companions than those without \24m flux. In a forthcoming paper (Oemler \etal, in preparation), we shall show that a large fraction of ICBS galaxies with very high values of SSFR are undergoing starbursts driven by galaxy interactions.

\subsection{Are Some Galaxies ``Young''?}

We have shown that starbursts are an unlikely explanation for the too-rapid evolution of galaxy luminosities and star formation rates. We have argued above that lowering the mean age of galaxies can explain the fast evolution of $M_B^*$, and Cowie \etal (1996) and Noeske \etal (2007b) have argued that the same change can explain the rapid evolution of the SSFR (Feulner \etal 2005b have raised the same possibility).   In this section we will examine this possibility, using the ICBS and local data samples. Figure 5a presents the cumulative distribution of SSFR for galaxies in the  Local data set, and  the $0.40 \le z < 0.60$ redshift range in the IR-ALL data set, down to a limiting mass $M_{gal} \ge 4\times10^{10}\Msun$, to which both data sets are essentially complete. We wish to compare the evolution shown in Fig 5a to the prediction of the exponential and delayed exponential models, as presented in Fig. 1, both  for an early time and for reduced ages of formation.

The fraction of objects with values of SSFR too high to explain by early--formation models is one indicator of young galaxies, but it is an insensitive test, since even quite young galaxies can have low values of SSFR if the SFR's fall from their peak values sufficiently rapidly.  Furthermore, this criterion depends on an absolutely correct calibration of star--formation rates, both observed and predicted by models. Instead, we will use the evolution rate of the SSFR distribution as our main tool. It is a characteristic of all of the models in Fig 1 that the curves of SSFR(t) diverge with time, and never cross. Therefore, given a set of galaxies at one epoch, ordered by SSFR, evolution with time will change the values of SSFR but will not change the rank order of the galaxies. Thus, the galaxies in the Local sample in Fig 5a, at any given value of $f$, should have evolved from the galaxies in the IR-ALL sample at the same value of $f$. Furthermore, if either set of models correctly describes the behavior of real galaxies, the value of $SSFR(t_o)$ at that value of $f$ will correspond to a unique $\tau$ model, from which we can predict the value of $SSFR(t)$ at the epoch of the earlier observations. There are two possible exceptions to these statements. Firstly, they ignore the growth of mass with time, which will cause some galaxies which are above the mass limit at the present epoch, to be below the mass limit, and excluded from the samples, at higher redshifts. However, we have simulated the effect of this on real data samples, and find that it has a negligible effect on the SSFR distribution. Secondly, they ignore the effect of mergers, which violate the assumption that a fixed population of objects evolves from one epoch to another. We will return to this issue shortly.

 We define $Z(SSFR(t_0)) = d(log(SSFR))/dt$, with units of $Gyr^{-1}$, and measure its value using the two data sets in Fig 5a, at multiple values of $f$ shown by red circles. In Figure 5b we compare the measured values of Z at these points with the predictions of the exponential and delayed exponential models from Fig 1, both for the early formation epochs (heavy solid and dashed lines) and for various delays in the formation epoch (lighter solid and dashed lines). Included are typical $1\sigma$ errors, calculated assuming random subsets drawn from an overall distribution of SSFR values with a shape like that of the observed galaxy samples. The results are inconsistent with a single epoch of formation for all galaxies with either set of star formation histories. However, at lower values of SSFR, beyond the 25th percentile (counting from high to low), the results lie close to the delayed exponential early--formation models. At higher values of SSFR, the data move steadily away from the heavy line, suggesting more and more recent epochs of formation. Thus, a reasonable conclusion is that most massive galaxies ($M_{gal}\ge 4\times10^{10}\Msun$) are coeval and old, but that about one quarter, those with values of SSFR greater than about $10^{-11}$ are ``younger'', with an epoch of formation which moves to later times with increasing SSFR.
 
   To examine the effect of mergers on this result, we perform a Monte Carlo simulation, using data from the IR-ALL sample in the redshift range $0.20 \le z < 0.40$. We take a merger rate versus redshift and galaxy mass as described in Xu \etal (2012) Equation 9. We assume that, after the merger, the mass is the sum of the mass of the two merging galaxies and that, after the merger is complete and any starburst has decayed, the total star formation rate is the sum of the pre--merger SFR's of the components. Although this is undoubtedly an oversimplification, a more elaborate treatment is not justified by our limited knowledge of the phenomenon. With these assumptions, the principal effect is a slight steepening of the SSFR distribution due to the averaging of the extremes of low and high SSFR values as pairs of galaxies combine. The effect on $Z$ is shown by the open circles in Fig. 5b. Except at very low values of SSFR (in which we are least interested) its amplitude is very modest and- given our ignorance of the details of the process- we shall ignore it in the remaining analysis.

 At this point we need to elaborate on what we mean by ``delayed formation''. The behavior of real galaxies which neither set of models can reproduce is the existence of  galaxies with both high and rapidly falling SSFR's. Exponential models peak at $t = t_{form}$; those with high values of SSFR at later times are those whose SSFR declines slowly. Delayed exponential models with arbitrarily high values of $\tau$ can peak at arbitrarily late times, but, like the exponential models, those with high values of SSFR at late times are those with slowly varying SSFR. It is obvious that {\em any} models (other than contrived cases seen at special epochs) which peak early must have either low values of SSFR or low values of $d(SSFR)/dt$ at late times. Only galaxies whose star formation histories peak at an age of the universe significantly greater than  the timescales on which star formation rises and falls before and after the peak can reproduce the behavior seen in our data. These we call ``young'' or ``delayed formation'' galaxies.  It must be emphasized that, by delayed formation, we are only talking about a delayed onset of the bulk of star formation. There is no way, with existing data, that one can rule out a tail of  star formation in all galaxies extending to quite early times. Thus, observations suggesting that there is some old stellar population in most if not all nearby galaxies (Grebel \& Gallagher 2004, Orban \etal 2008  ) does not contradict our picture.
 
Cowie \etal (1996) and Noeske \etal (2007) observe downsizing: not only are some galaxies young, but their age decreases with decreasing mass.  To examine this effect, we divide our samples by mass. Figure 6 repeats the analysis in Fig. 5, but dividing galaxies into 3 mass ranges, $4\times10^{10}\Msun-7\times10^{10}\Msun$,  $7\times10^{10}\Msun-1.4\times10^{11}\Msun$, and $ M_{gal}>1.4\times10^{11}\Msun$. There is a clear divide, at approximately $M_{gal} = 1.4\times10^{11}\Msun$. Galaxies above this limit follow the old delayed exponential locus; but at least some galaxies below this limit need a range of younger ages. Galaxies in the mass range $7\times10^{10}\Msun-1.4\times10^{11}\Msun$ diverge from the old locus below the 30th percentile, galaxies in the $4\times10^{10}\Msun-7\times10^{10}\Msun$  mass range depart from the old locus below the 40th percentile, a very large fraction of young galaxies, indeed. Note, however, that at every mass less than $M_{gal} = 1.4\times10^{11}\Msun$ there is a large range of age, with a majority of old, coeval galaxies. Although mean age is a function of mass, such an averaging hides a more complicated story.

%Figure 6
\begin{figure}[h]
\figurenum{6}
\epsscale{1.1}
\plotone{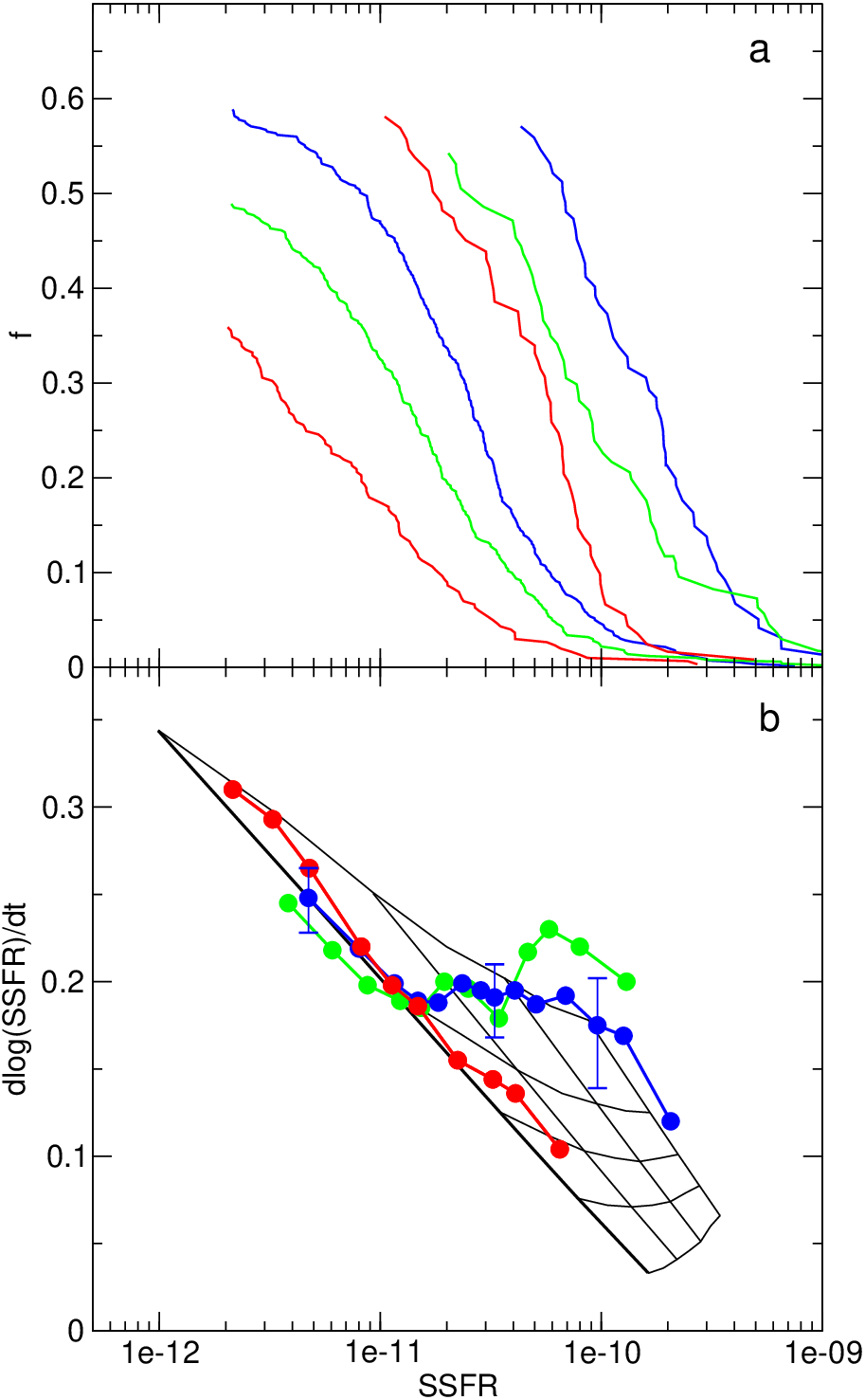}
\figcaption{ a- cumulative distribution of specific star formation rate in 3 mass ranges. Left 3 curves- PM2GC local sample; right 3 curves- IR-ALL sample $0.4 \le z < 0.6$. Blue curve- $4\times10^{10}\Msun<M<7\times10^{10}\Msun$,  green curve- $7\times10^{10}\Msun<M<1.4\times10^{11}\Msun$, red curve-  $ M_{gal}>1.4\times10^{11}\Msun$ b- evolutionary rate of SSFR vs $SSFR(t_o)$. Colors correspond to mass ranges in Fig 6a; the grid of models is the delayed exponential grid from Fig. 5b\vspace{1.0cm}}
\end{figure}

 The expected errors in this case are more significant than those shown in Figure 5b. Evaluating the overall significance of the departure of, for example, the red and blue curves from these typical errors is not straightforward, because the curves are derived from cumulative distributions and the errors of adjacent points are, therefore, at least partially correlated. Instead, we perform Monte Carlo simulations on pairs of distributions, drawn from the same parent population, and of the size of our samples, and ask how often the two resulting  Z curves depart from each other, at every value of $f$, by as much as the pairs of Z curves in Fig 6b. We find that, for example, the chance that the red and blue curves were drawn from the same population is only $2\times10^{-5}$, so the observed differences are very significant.

%Figure 7
\begin{figure}[h]
\epsscale{1.1}
\figurenum{7}
\plotone{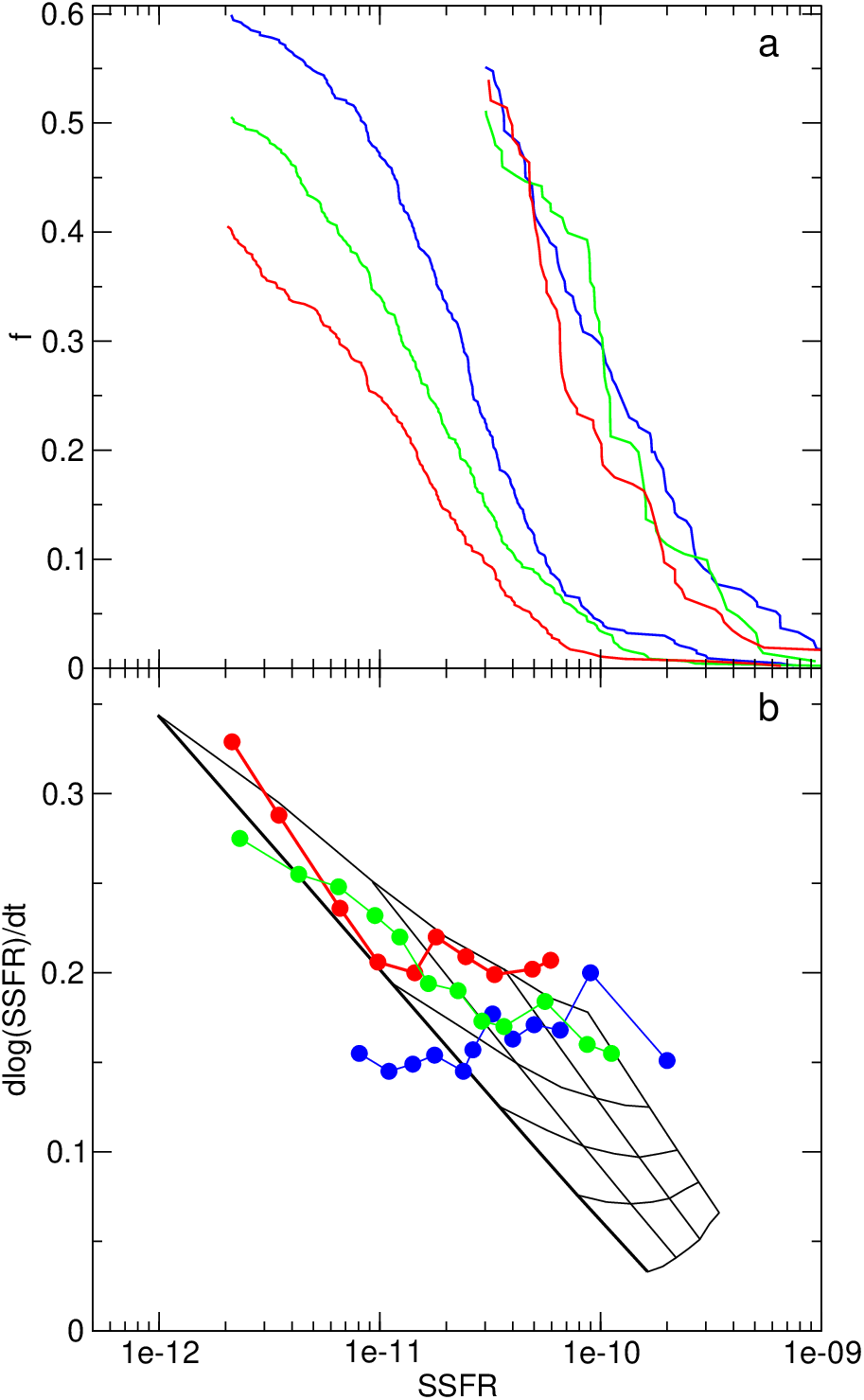}
\figcaption{Same analysis as in Fig. 6, but with galaxies from the PM2GC local sample, and the 4-Fields $0.4 \le z < 0.6$ sample, divided into intervals of local density. Blue- lowest third values of $\rho/\rho_{med}$, Green- middle third values  of $\rho/\rho_{med}$, Red- highest third values of  $\rho/\rho_{med}$}
\end{figure}

One might expect that galaxy formation and evolution would proceed more rapidly in high--density regions of the universe than in low--density regions (e.g. Cen 2011). Therefore, in Figure 7, we test for dependance of galaxy evolution on local density which we calculate from the distance to the third nearest neighbor with mass $M_{gal} \ge 4 \times10^{10}\Msun$. We divide the distribution of $\rho/\rho_{med}$, where $\rho_{med}$ is the median value of $\rho$ for galaxies in the same redshift interval, into three parts, and examine the evolution of SSFR for each third. The nearby sample is the PM2GC sample, and the distant sample consists of galaxies in the 4Fields sample with $0.4 \le z < 0.6$. Because, as we have shown, the evolution of SSFR depends on galaxy mass, and because the distribution of mass is dependent on local density (Vulcani \etal 2012a) we weight galaxies in each density interval by mass to produce weighted distributions with mass which are the same in all intervals. Although the behavior of galaxies in the lowest third of the density distribution differs a bit from the others, all 3 groups seem to display the same wide range in formation epochs, and thus provide no evidence that the age distribution of galaxy populations depends on local density.

%Figure 8
\begin{figure}[h]
\epsscale{1.1}
\figurenum{8}
\plotone{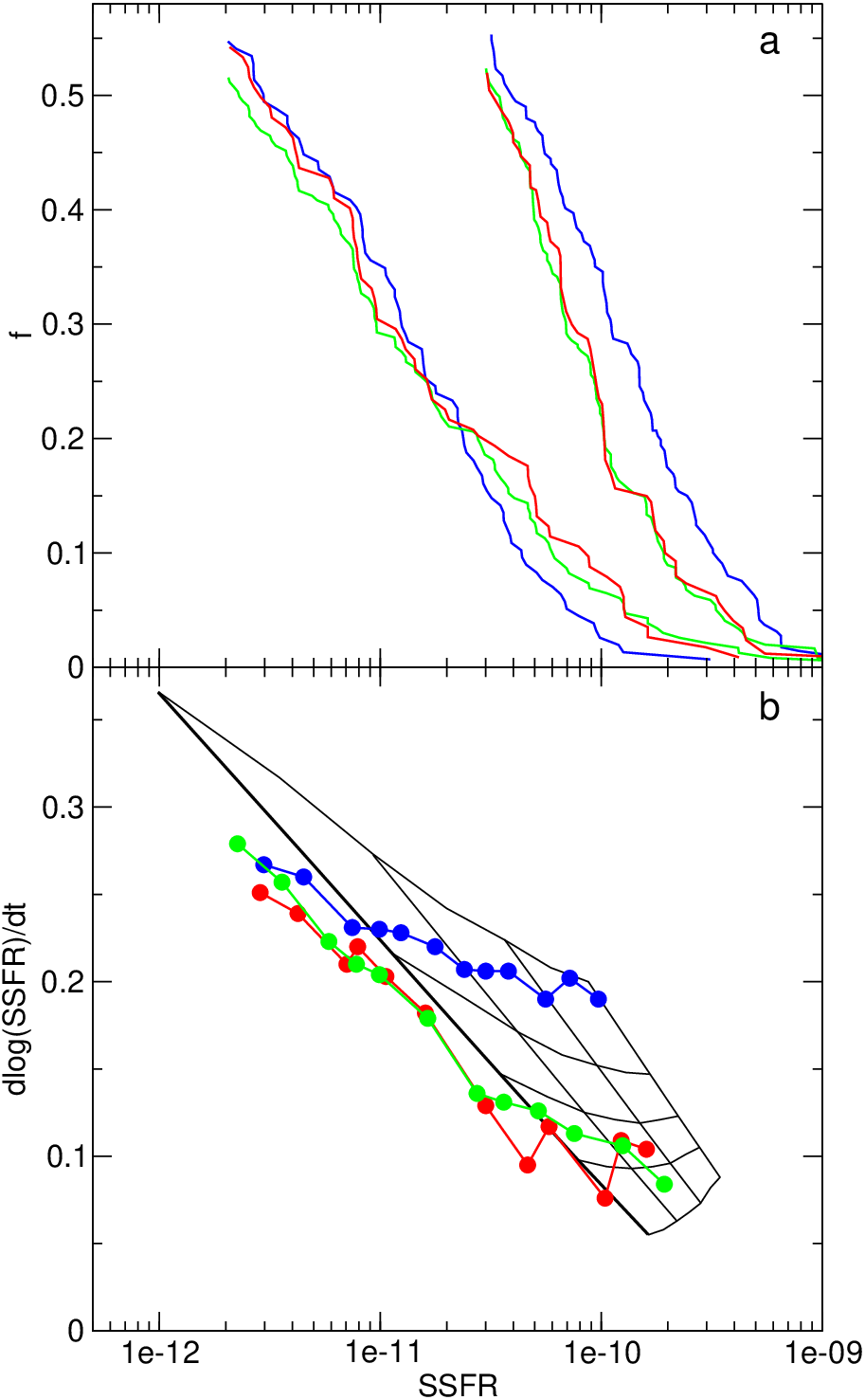}
\figcaption{Same analysis as in Fig. 6, but with galaxies from the NGP local sample, and the 4Fields $0.4 \le z < 0.6$ sample, divided between isolated galaxies (blue), all group members (green), and members of groups with 4 or more members (red)}
\end{figure}

 Figure 8 presents the behavior of the same redshift intervals analyzed in Figures 6 and 7, now divided by group membership.  We have constructed a group catalog, for the redshift interval $0.2 \le z < 0.6$, from the 4 Fields sample  using the standard Huchra \& Geller (1982) friends--of--friends method. Galaxies are linked if separated by a projected distance on the sky, $D \le D_{lim}$, and a velocity difference $\delta v \le \delta v_{lim}$. In order to minimize redshift dependences, we use a sample with a fixed limiting value  $M-M^* \le 2.0$. We take $\delta v_{lim} = 350 km s^{-1}$ and $D_{lim} = 520C^{-1/2}$kpc, where $C$ is the completeness of the spectroscopic sample as a function of r magnitude and field position. With these selection parameters, about half of the galaxies are assigned to groups, and half are isolated. For a local comparison, we have constructed a group catalog from the NGP sample using the same selection criteria (the PM2GC sample is of a very narrow strip, making the identification of groups problematic). Although the mass distribution varies less with group membership than with local density (Vulcani \etal 2012b) we still weight galaxies in the two groups to produce similar mass distributions. In contrast to Figure 7, we see here a clear difference in behavior with environment: only isolated galaxies appear to be young, galaxies with at least one massive neighbor follow the evolution expected for an early formation epoch.

\section{Discussion}

The analysis of the previous section indicates  the unexpectedly rapid evolution of star formation cannot be explained by starbursts, but can be understood as a consequence of the delayed formation of some fraction of less massive galaxies. However, this analysis has not provided much quantitative detail on the age distribution of galaxies. Given the infinite variety of possible forms of star formation histories, the limited information- distributions of SSFR's- we have used is insufficient to do so. Noeske \etal (2007b) present one possible scenario- exponentially declining Bruzual \& Charlot models, like those presented in Fig.1, with a delay in formation redshift of the form $(1+z_{form}) \sim M_{gal}^\beta$, but it is hardly unique. It is also clearly not correct, since Fig. 6 shows that, at every mass, many galaxies appear to be old, and only the fraction of younger objects changes with mass, and Figure 8 suggests that only isolated galaxies may be younger. A more complete description of galaxy formation is not possible within the limitations of the data we have considered.

There is, however, more information available. The entire 2--dimensional distribution of SSFR-Mass values of galaxies over the redshift range $0.0 \le z < 1.0$, plus the integrated star formation history of galaxies back the redshifts of order 3, which has been studied by many workers (see Cucciati \etal 2012 for references) provides additional constraints. In addition, if one makes the common assumption that the different structures of bulges and disks reflect different timescales for star formation in these subsystems, the distribution of bulge--to--disk ratios of galaxies provide an additional constraint. In Paper IV we will make a start at such a complete analysis: using the distribution of SSFR vs mass and redshift, as well as the history of the specific star formation density with redshift to constrain the evolution of galaxies with log--normal star formation histories. We will, therefore, defer all further discussion of this topic to the following paper.

One item which we will not address in Paper IV is the environmental dependance of galaxy ages. Figures 7 and 8 seem to be telling different stories: galaxy age depends on group membership but not on local galaxy density. These results are not, however, as inconsistent as they appear. Although  the two measures of environment are correlated, the correlation is very broad:  almost half of the isolated galaxies fall into the highest two thirds of relative density, and 20 percent of the grouped galaxies fall into the lowest third of relative density. In general, our measure of local density samples a significantly larger volume than does our group/isolated criterion. This suggests that it is the very local environment which affects galaxy age: a not implausible conclusion. Together, the dependance on galaxy mass and group environment suggest that galaxy age might be dependent on only a single physical parameter: total mass within a volume of size  about 0.5Mpc.

 The more sophisticated theoretical expectations are less than clear. The models of Cen (2011) produce more rapid downsizing in denser regions, but Cen does not distinguish between archaeological downsizing and downsizing in time (to use the terminology of Neistein \etal 2006). Neistein \etal\_ state that downsizing in time, the mass-dependent delay in galaxy formation discovered by Cowie \etal (1995) cannot be explained without invoking poorly understood hydrodynamic processes, an opinion shared by Fontanot \etal (2009), so the concordance of theory and observations is not yet established.

\section{Acknowlegements}

We are grateful to Kai Noeske for providing us with the AEGIS data. Oemler and Dressler acknowledge support of the NSF grant AST-0407343. This work is based in part on observations made with the Spitzer Space Telescope, which is operated by the Jet Propulsion Laboratory, California Institute of Technology under a contract with NASA. Support for this work was provided to Dressler and Oemler by NASA through an award issued by JPL/Caltech.  Vulcani and Poggianti acknowledge financial support from ASI contract I/016/07/0 and ASI-INAF I/009/10/0. This research has made use of the NASA/IPAC Extragalactic Database (NED) which is operated by the Jet Propulsion Laboratory, California Institute of Technology, under contract with the National Aeronautics and Space Administration. We have made extensive use of Ned WrightÕs Online Cosmological Calculator (Wright 2006).

\clearpage

\end{document}